**Lexicographic Preferences over Random Availability Functions**

**By**

**Somdeb Lahiri**

**ORCID: https://orcid.org/0000-0002-5247-3497**

**(Formerly) PD Energy University (EU-G)**

**(somdeb.lahiri@gmail.com)**

**May 15, 2025.**

**Abstract**

We provide an axiomatic characterization of lexicographic preferences over the set of all random availability functions using two assumptions. The first assumption is strong monotonicity, which in our framework is equivalent to the strong dominance property in microeconomics. The second assumption is independence of worse alternatives and we show that a weaker version of the same suffices for our purpose.

**1. Introduction:** Consider a person who on a working day has to choose between two wagons to buy a sandwich for lunch. The two restaurants, while reasonably close to his office are in opposite directions, so that it is not possible for him to visit both restaurants during the relatively short lunch break. Suppose that the sandwiches on the menu card of the two restaurants are the same and come in two varieties: fish and chicken. However, due to lunch hour rush, it is quite possible that by the time he arrives at any one of the two wagons, one or both of the two types of sandwiches would be "sold-out". Hence there is some probabilistic uncertainty about the availability of each type of sandwich at the wagons and these probabilities may be different for different wagons. The person's problem is to choose his preferred destination.

This kind of decision-making problem is one example of the theoretical model discussed in Lahiri (2025), where the object of choice- which in the case of the situation discussed in the previous paragraph is a wagon- is referred to as a "random availability function". In Lahiri (2025), the main result is a proposition which provides two plausible conditions which together are sufficient for preferences over random availability functions to have a numerical representation. Of the two assumptions required for the proposition, the second assumption ("Weak Continuity") is a technical one, and the plausibility of a technical assumption, is more often determined by "academic conditioning" than by unbiased reasoning and this may explain why the only example of preferences over random availability provided in Lahiri (2025), has a numerical representation, although the preference reflects "incomplete" reasoning on the part of the decision maker. This in no way detracts any merit from the main result in Lahiri (2025) or the example, the purpose of the example being merely to show that the proposition is not vacuously true. There are several examples of numerical

representations- notably a weighted sum of the logarithm of the individual probabilities- that may be meaningful representations of preferences over random availability functions.

In this note, we propose "lexicographic preferences" as a very acceptable preference over the set of all random availability functions. It is well known that such preferences do not allow numerical representation (see example 3.2 in page 83 of Varian (1978)).

The first known axiomatic characterization of lexicographic preferences in an abstract framework is available in Fishburn (1975). This line of research has been pursued in the work of Petri and Voorneveld (2016). Axiomatic characterizations of lexicographic preferences for the model of classical consumer demand theory are available in Mitra and Sen (2014) and Goswami, Mitra and Sen (2022). The difference between the work of Goswami, Mitra and Sen (2022) and the others is that the former invokes continuity assumptions (and hence topological considerations), whereas the others rely totally on combinatorial methods, and it is precisely because of this, that the work in Mitra and Sen (2014) is an alternative proof of the axiomatic characterization in Fishburn (1975), which the former claims to be.

We provide an axiomatic characterization of lexicographic preferences using combinatorial methods. We use two assumptions to characterize lexicographic preferences in our framework. The first is a stronger version of weak dominance and the second is similar in spirit to what Petri and Voorneveld (2016) refer to as non-compensation. Non-compensation originates in the work of Fishburn (1975). We refer to our second assumption as "independence of worse alternatives" although as we show in our main result, a weaker version of the assumption suffices for our purpose. The weaker version makes its application in the proof, somewhat easier to guess.

Our assumptions are few, easy to understand and allow for a very easy proof of the characterization of lexicographic preferences.

**2. The framework of analysis:** Our notations are exactly as in Lahiri (2025).

Let $\mathbb{N}$ denote the set of natural numbers, $\mathbb{R}$ the set of real numbers and $\mathbb{R}_+$ the set of non-negative real numbers. Let $[0, 1] = \{\alpha \in \mathbb{R} | 0 \leq \alpha \leq 1\}$.

Let X be a non-empty finite set of alternatives containing at least two alternatives.

A "**random availability function" (RAF)** (on X) is a function A: X$\rightarrow$ [0, 1], such that for each x$\in$X, A(x) is the probability that x is available and 1-A(x) is the probability that x is not available.

Let $\mathcal{P}$(X) denote the set of all RAF's on X. Thus $\mathcal{P}(X) = [0, 1]^X$.

Let $\succcurlyeq$ be binary relation on $\mathcal{P}$(X). Instead of writing (A, B)$\in \succcurlyeq$, where A, B$\in \mathcal{P}$(X), to economize on notation we will write A $\succcurlyeq$ B.

We assume that $\succcurlyeq$ is reflexive (i.e., A $\succcurlyeq$ A for all A$\in\mathcal{P}$(X)), connected (i.e., for all A, B$\in\mathcal{P}$(X) with A $\neq$ B, either A $\succcurlyeq$ B or B $\succcurlyeq$ A) and transitive (i.e., for all A, B, C$\in\mathcal{P}$(X): [A $\succcurlyeq$ B and B $\succcurlyeq$ C] implies [A $\succcurlyeq$ C].

Such a binary relation $\succcurlyeq$ is called a **preference relation** with the interpretation that if for A, B$\in\mathcal{P}$(X), it is the case that A $\succcurlyeq$ B, then A is "**at least as good as**" B.

The asymmetric part of $\succcurlyeq$, denoted by $\succ$ is interpreted as "**is strictly preferred to**" and the symmetric part of $\succcurlyeq$, denoted by $\sim$ is interpreted as "**no different from**".

Given RAF's A and B, A is said to **strictly dominate** B, if A(x) > B(x) for all $x \in X$.

$\succcurlyeq$ is said to be "**weakly dominant**" if for A, B$\in \mathcal{P}(X)$: [A strictly dominates B] implies [A $\succ$ B].

A function u: $\mathcal{P}(X) \rightarrow \mathbb{R}$ is said to be a **utility function for (numerical representation of)** a preference relation $\succcurlyeq$ if for all A, B$\in \mathcal{P}(X)$: [A$\succcurlyeq$B] <u>if and only if</u> [u(A) $\geq$ u(B)].

**3. Example in Lahiri (2025):** Let $\pi$: $X \rightarrow \mathbb{R}_+$ be a function such that for all $x \in X$, $\pi(x)$ is the pay-off from choosing x. Thus, if $x \in X$ is the desired alternative, then the expected pay-off from A is $\pi(x)A(x)$. Let u:$\mathcal{P}(X) \rightarrow \mathbb{R}$ be such that for all A$\in \mathcal{P}(X)$, $u(A) = \max_{x \in X} \pi(x)A(x)$ and $\succcurlyeq$ be the preference relation on $\mathcal{P}(X)$ such that for all A, B$\in \mathcal{P}(X)$: A $\succcurlyeq$ B if and only if u(A) $\geq$ u(B). Clearly, u is a utility function for $\succcurlyeq$.

We may refer to the preference relation defined in the previous paragraph as **maximum expected pay-off relation**.

One of the problems with the maximum expected pay-off relation is the following.

Suppose, X = {\$40, \$10}, A, B$\in \mathcal{P}(X)$, with A(10) = $\frac{4}{5}$ (i.e., 80%), A(40) = $\frac{1}{5}$ (i.e., 20%), B(10) = $\frac{9}{10}$ (i.e., 90%) and B(40) = $\frac{1}{10}$ (i.e., 10%).

It is worth emphasizing that {\$10} and {\$40} are <u>not</u> necessarily mutually exclusive events although the probabilities in both RAF's add up to one. In fact, the two events are mutually exclusive <u>if and only if</u> the probability of the event "\$10 or \$40" is 1. There is no such assumption in this context.

Suppose $\pi(10) = 10$ and $\pi(40) = 40$.

Then according to the maximum expected pay-off relation B is strictly preferred to A, although the higher pay-off is twice as likely at A than at B, and the lower pay-off is 10% more likely at B than at A. Further, there is considerable difference between the pay-offs. Looking at the numbers, it is possible that many might prefer A to B.

The second problem, with the maximum expected pay-off relation is that after having chosen B, the decision maker would naturally prefer \$40 to \$10. What happens if \$40 is not available after RAF B is chosen? In that case, the decision-maker is left with the choice of realizing \$10 with 90% chance. In such a situation it is quite possible that the decision-maker would regret having foregone a choice between a 20% chance of getting \$40 and 80% chance of getting \$10 and perhaps have available both the outcomes to choose from, had he chosen A.

In order to avoid such regret, decision based on lexicographic preferences may prove useful.

**4. Lexicographic preferences:** Suppose for the sake of simplicity R is a linear order (reflexive, connected, transitive and anti-symmetric binary relation) on X, where for some positive integer K, X is a set of K distinct alternatives $\{x_1, \ldots, x_K\}$. Without loss of generality suppose that for all i, j$\in$ {1, …, K} with $i \neq j$: [i < j] implies [$x_i$ R $x_j$ but <u>not</u> $x_j$ R $x_i$].

The **lexicographic preference relation** on $\mathcal{P}$(X) denoted by $\succcurlyeq^{Lex}$ is defined as follows: given A, B$\in$ $\mathcal{P}$(X), A$\succcurlyeq^{Lex}$B if and only if either (i) A = B or (ii) A $\neq$ B and $A(x_k) > B(x_k)$, where $k = \min\{i \mid A(x_i) \neq B(x_i)\}$.

Hence, when the above conditions are not fulfilled, we have B $\succcurlyeq^{Lex}$A and not A$\succcurlyeq^{Lex}$B.

Let $\succ^{Lex}$ denote the asymmetric part of $\succcurlyeq^{Lex}$ and let $\sim^{Lex}$ denote the symmetric part of $\succcurlyeq^{Lex}$.

It is easy to see that $\succcurlyeq^{Lex}$ is a linear order on $\mathcal{P}$(X).

In the context of the numerical example, since A(40) = $\frac{1}{5} > \frac{1}{10}$ = B(40), it must be that according to lexicographic preference A is strictly preferred to B.

In general, $\succcurlyeq^{Lex}$ satisfies a stronger version of weak dominance criteria.

$\succcurlyeq$ is said to be **strongly monotonic** if for A, B$\in\mathcal{P}$(X) and x$\in$X: [A(y) = B(y) for all y$\in$X\{x} and A(x) > B(x)] implies [A $\succ$ B].

Given our definition of preferences, by successive application of the strongly monotonic assumption, we arrive at the weak dominance criteria.

An equivalent way of defining strongly monotonic preferences in our framework is the following:

$\succcurlyeq$ is strongly monotonic if and only if for all A, B$\in\mathcal{P}$(X) with A $\neq$ B: [A(x) $\geq$ B(x) for all x$\in$X] implies [A $\succ$ B].

This latter formulation is usually deferred to as "strong dominance".

Given our definition of preferences, by successive application of the strongly monotonic assumption, we arrive at strong dominance.

Strongly monotonic is the consequence of Lemma 2.1 in Petri and Voorneveld (2016).

The following is a property for preference relations that is available in Petri and Voorneveld (2016). It originates in Fishburn (1975) and is Axiom 3 in Mitra and Sen (2014).

$\succcurlyeq$ is said to satisfy **non-compensation** if for all A, B, C, D$\in\mathcal{P}$(X): [{x$\in$X| A(x) > B(x)} = {x$\in$X| C(x) > D(x)} and {x$\in$X| A(x) < B(x)} = {x$\in$X| C(x) < D(x)}] implies [A $\succcurlyeq$ B if and only if C $\succcurlyeq$ D].

Non-compensation implies Axiom 2 in Mitra and Sen (2014), which may be stated thus:

For all A, B, C, D$\in\mathcal{P}$(X) and y$\in$X: [A $\neq$ B, A(x) = B(x) for all x$\in$X \{y}, C(x) = D(x) for all x$\in$X \{y}, A(y) = C(y) and B(y) = D(y)] implies [A $\succcurlyeq$ B if and only if C $\succcurlyeq$ D].

The reason for this implication is {x$\in$X\{y}| A(x) > B(x)} = {x$\in$X\(y)| C(x) > D(x)} = $\phi$ and {x$\in$X\{y}| A(x) < B(x)} = {x$\in$X\{y}| C(x) < D(x)} = $\phi$.

We now introduce an assumption similar in nature to non-compensation.

$\succcurlyeq$ is said to satisfy **independence of worse alternatives** if for all A, B, C, D$\in\mathcal{P}$(X) and k$\in${1,…, K}: [$A(x_k) \neq B(x_k)$, $\{i \leq k \mid A(x_i) < B(x_i)\} = \{i \leq k \mid C(x_i) < D(x_i)\}$ and $\{i \leq k \mid A(x_i) > B(x_i)\} = \{i \leq k \mid C(x_i) > D(x_i)\}$] implies [A $\succcurlyeq$ B if and only if C $\succcurlyeq$ D].

It is easily verifiable that $\succcurlyeq^{Lex}$ satisfies independence of worse alternatives.

For our purposes a weaker version of independence of worse alternatives would suffice.

$\succcurlyeq$ is said to satisfy **weak independence of worse alternatives** if for all A, B, C, D$\in \mathcal{P}(X)$ such that $A \neq B$: [$k = \min \{i \in \{1,\ldots, K\} \mid A(x_i) \neq B(x_i)\} = \min \{i \in \{1,\ldots, K\} \mid C(x_i) \neq D(x_i)\}$, $(A(x_k) - B(x_k))(C(x_k) - D(x_k)) > 0$] implies [$A \succcurlyeq B$ if and only if $C \succcurlyeq D$].

Note that if $k = \min \{i \in \{1,\ldots, K\} \mid A(x_i) \neq B(x_i)\} = \min \{i \in \{1,\ldots, K\} \mid C(x_i) \neq D(x_i)\}$ and $k > 1$, then $A(x_k) \neq B(x_k)$, $\{i < k \mid A(x_i) < B(x_i)\} = \{i < k \mid C(x_i) < D(x_i)\} = \phi$ and $\{i < k \mid A(x_i) > B(x_i)\} = \{i < k \mid C(x_i) > D(x_i)\} = \phi$.

Thus, independence of worse alternatives implies weak independence of worse alternatives.

**5. Axiomatic characterization of lexicographic preferences:** We now state and prove the main characterization result for lexicographic preferences.

**Proposition 1:** $\succcurlyeq$ is strongly monotonic and satisfies weak independence of worse alternatives if and only if $\succcurlyeq \,=\, \succcurlyeq^{Lex}$.

**Proof:** Since it is easy to verify that $\succcurlyeq^{Lex}$ satisfies the two properties, let us suppose $\succcurlyeq$ is strongly monotonic and satisfies weak independence of worse alternatives.

Let A, B$\in \mathcal{P}(X)$ be such that $A \neq B$.

Let $k = \min \{i \in \{1,\ldots, K\} \mid A(x_i) \neq B(x_i)\}$. Thus, if $k > 1$, then $A(x_i) = B(x_i)$ for all $i = 1, \ldots, k-1$.

Let C$\in \mathcal{P}(X)$ be such that $C(x_i) = B(x_i)$ for all $i \leq k$ and if $k < K$, then $C(x_i) = A(x_i)$ for all $i > k$.

Thus, $C(x_i) = A(x_i)$ for all $i \neq k$ and $C(x_k) \neq A(x_k)$.

Since $\succcurlyeq$ is strongly monotonic, $C \succ A$ if $C(x_k) > A(x_k)$ and $A \succ C$ if $A(x_k) > C(x_k)$.

However, $C(x_i) = B(x_i)$ for all $i \leq k$ implies $k = \min \{i \in \{1,\ldots, K\} \mid A(x_i) \neq B(x_i)\} = \min \{i \in \{1,\ldots, K\} \mid A(x_i) \neq C(x_i)\}$, $(A(x_k) - B(x_k))(A(x_k) - C(x_k)) > 0$].

Thus, by weak independence of worse alternatives $A \succcurlyeq B$ if and only if $A \succcurlyeq C$.

Hence, $C \succ A$ implies $B \succ A$ and $A \succ C$ implies $A \succ B$.

Since $C(x_k) = B(x_k)$, we get $B \succ A$ if $B(x_k) > A(x_k)$ and $A \succ B$ if $A(x_k) > B(x_k)$.

Thus, $\succcurlyeq \,=\, \succcurlyeq^{Lex}$. Q.E.D.

**Corollary of Proposition 1:** $\succcurlyeq$ is strongly monotonic and satisfies independence of worse alternatives if and only if $\succcurlyeq \,=\, \succcurlyeq^{Lex}$.

**Acknowledgment:** I wish to thank Subhadip Chakrabarti for very useful comments and suggestions for improvement on an earlier version of this paper.

**References**

1. Fishburn, P.C. (1975): Axioms for lexicographic preferences. Review of Economic Studies, Volume 42, Number 3, Pages 415-419.

2. Goswami, M.P., Mitra, M. and Sen, D. (2022): A characterization of lexicographic preferences. Decision Analysis, Volume 19, Number 2, Pages 170-187.

3. Lahiri, S. (2025): Numerical Representation of Preferences over Random Availability Functions: A Note. https://doi.org/10.48550/arXiv.2504.18863

4. Mitra, M. and Sen, D. (2014): An alternative proof of Fishburn's axiomatization of lexicographic preferences. Economics Letters, Volume 124, Number 2, Pages 168-170.

5. Petri, H. and Voorneveld, M. (2016): Characterizing lexicographic preferences. Journal of Mathematical Economics, Volume 63, Issue C, Pages 54 -61.

6. Varian, H. (1978): "Microeconomic Analysis", W.W. Norton and Company, INC.